\documentclass[twocolumn,superscriptaddress,showpacs,amsmath,amssymb,prb,longbibliography]{revtex4-1}

\usepackage[pdftex]{graphicx} 
\usepackage{dcolumn}  
\usepackage{bm}       
\usepackage[usenames,dvipsnames]{color}
\definecolor{URLCOL}{rgb}{0,0.52,0.83} 
\definecolor{LINKCOL}{rgb}{0.05,0.5,0} 
\definecolor{orange}{rgb}{0.6,0.3,0} 
\definecolor{CITECOL}{rgb}{0.25,0,0.48} 
\usepackage{epstopdf}

\definecolor{dkgreen}{rgb}{0,0.5,0}

\usepackage[normalem]{ulem}

\usepackage{booktabs}
\usepackage{natbib}
\usepackage{tabularx}
\usepackage{rotating}

\def\ext{_{\rm ext}}

\graphicspath{{f/}}

\definecolor{TITLECOL}{rgb}{0.1,0.2,0.7} 
\definecolor{SECOL}{rgb}{0.1,0.2,0.7} 
\definecolor{CONTENTSCOL}{rgb}{0.1,0.2,0.7} 
\definecolor{SSECOL}{rgb}{0.25,0,0.48} 
\definecolor{SSSECOL}{rgb}{0.2,0.08,0.53} 
\definecolor{FINCOL}{rgb}{0.01,0.3,0.07} 

\definecolor{URLCOL}{rgb}{0,0.17,0.43} 
\definecolor{LINKCOL}{rgb}{0.05,0.4,0} 
\definecolor{CITECOL}{rgb}{0.35,0,0.48} 

\def\bea{\begin{eqnarray}}
\def\eea{\end{eqnarray}}
\def\ben{\begin{equation}}
\def\een{\end{equation}}
\def\benu{\begin{enumerate}}
\def\enu{\end{enumerate}}

\def\bei{\begin{itemize}}
\def\eei{\end{itemize}}
\def\beit{\begin{itemize}}
\def\eit{\end{itemize}}
\def\benu{\begin{enumerate}}
\def\enu{\end{enumerate}}

\def\br{{\bf r}}

\def\ee{_{\rm ee}}

\def\jindex{N_\mathrm{occ}}
\def\a{a} 

\def\br{\mathbf{r}{}} 

\begin{document}

\title{Accurate correlation energies in one-dimensional systems\\ from small, system-adapted basis functions}
\author{Thomas E.~Baker}
\affiliation{Department of Physics \& Astronomy, University of California, Irvine, California 92697 USA}
\author{Kieron Burke}
\affiliation{Department of Chemistry, University of California, Irvine, California 92697 USA}
\affiliation{Department of Physics \& Astronomy, University of California, Irvine, California 92697 USA}
\author{Steven R.~White}
\affiliation{Department of Physics \& Astronomy, University of California, Irvine, California 92697 USA}
\date{\today}

\begin{abstract}
We propose a general method for constructing 
system-dependent basis functions for correlated quantum calculations.  Our construction combines features from several
traditional approaches: plane waves, localized basis functions,
and wavelets.  In a one-dimensional mimic of Coulomb systems,
it requires only 2-3 basis functions per electron to achieve high accuracy,
and reproduces the natural orbitals.  We illustrate its
effectiveness for molecular energy curves and chains of many one-dimensional atoms.
We discuss the promise and challenges for realistic quantum chemical
calculations.
\end{abstract}

\maketitle

\section{Introduction}
\label{intro}

Many tens of thousands of electronic structure calculations are performed each
year, the vast majority in a single-particle basis set of some sort.  
These calculations can be divided into two types: those that extract the 
energy from a set of single-particle occupied
orbitals (denoted single-determinant) such as  density
functional theory (DFT) \cite{HK64,KS65,BW13,B12,PGB14} or Hartree Fock (HF), 
and those that go beyond a single determinant, such as configuration interaction, \cite{sherrill1999configuration,cramer2013essentials} 
coupled cluster methods, \cite{coester1960short,cizek1980coupled,vcivzek1966correlation}
density matrix renormalization group (DMRG), \cite{W92,W93,WM99,chan2002highly,U05,U11} and some
types of quantum Monte Carlo.  
Going beyond a single determinant is necessary for many systems, but is
typically much more
demanding computationally.  
Such calculations are more
difficult because larger basis sets are needed
to achieve chemical accuracy (1 kcal/mol), and
computation times usually scale as a high power of the number of
basis functions.  
These larger
basis sets are needed to represent the electron-electron cusp in the
wavefunction which exists at every point in space.

A natural question arises:  what would be the optimal basis set for an
electronic structure calculation, assuming the basis is specifically adapted to
that system? 
For a single-determinant method, the answer is clear:  the self-consistent
occupied orbitals 
are the optimal basis for that calculation: used as a basis, they
reproduce the exact energy and properties.  
The number of these basis functions (for a spin-restricted
calculation) is thus $N_e/2$, where $N_e$ is the number of electrons.  
Of course, this minimal basis does not offer a computational shortcut:  the occupied
orbitals must be determined in a separate, non-adapted basis calculation.
Here, we are concerned with multi-determinant methods, and we will assume that the
computation time for a traditional single-determinant calculation is small in comparison to
the multi-determinant method.

For post-HF methods, there is {\it no} exact finite system-adapted basis:
any finite basis introduces errors.  
However, the natural orbitals are close to the most rapidly converging
single-particle basis, at least in terms of allowing the greatest possible
overlap with the exact ground state.\cite{lowdin1956natural,lowdin1955quantum}
The natural orbitals are the eigenstates of the single-particle density matrix
(also known as the equal-time one-particle Green's function). 
The number of nonzero eigenvalues (occupancies) is infinite. 
A (near) optimal basis of $M_{no}$ orbitals
consists of the $M_{no}$ natural orbitals with the greatest occupancy.

One obvious weakness in using natural orbitals is that one does not know them until
after one has solved the interacting system, using a post-HF method, with another
larger basis.
Iterative natural orbital methods are a way to reduce the computational expense,
but {\it approximate}  natural orbitals that did not need a post-HF method to
determine them could be very useful.\cite{jensen1988second}  
But natural orbitals have another key weakness:  they are (normally)
completely delocalized across the system.  
This delocalization prevents a number of shortcuts that can greatly decrease
computation times for large systems. 
Delocalization is especially harmful for low-entanglement methods such as DMRG,
since there is no area law for the entanglement entropy in a delocalized basis.\cite{eisert2010colloquium}

Here we describe an approach that starts with the occupied orbitals of
a DFT (or HF) calculation, and yields
basis sets which produce high accuracy in correlated calculations.
We test this approach in 1D, using potentials that make 1D mimic 3D in many
respects, and using DMRG.\cite{BSWB15}   
The computational effort for the basis construction is minimal.  
The number of basis functions needed is typically about $2M_{no}$, where $M_{no}$ is
the minimal number of natural orbitals needed to reach
high accuracy, or about two or three times the number of electrons.
We expect this method can be easily extended to quasi-1D systems
(such as large-$Z$ atoms or chains of real H atoms) and hope it can be
applied more generally in 3D.

The first step produces what we call ``product plane waves'' (PPWs)
by multiplying the occupied
orbitals by a set of low momentum cutoff plane waves.  
The lowest momentum is determined
by the spatial extent of the entire system.
This simple ansatz converges well in our tests in 1D, and we
show how its convergence is within about a factor of 2 compared
to natural orbitals. 
But a weakness of PPWs, shared with natural orbitals, is that the basis is not local.
As the second major part of this work,
we describe fragmentations of the PPWs that utilize wavelets\cite{haar1910theorie,gabor1946theory,grossmann1984decomposition,meyer1989orthonormal,mallat1989theory,daubechies1992ten,daubechies1988orthonormal,daubechies1993orthonormal,wei1996wavelets,tymczak1997orthonormal,harrison2004multiresolution,fann2005mra,harrison2005multiresolution,fann2007madness,thornton2009introducing,harrison2016madness,van2004wavelets,natarajan2011wavelets,PhysRevA.76.040503,flores1993high,flad2002wavelet,bischoff2012computing,bischoff2013computing,beylkin2008approximating,genovese2011daubechies,natarajan2011wavelets,arias1999multiresolution,harrison2005multiresolution,khoromskij2011numerical,nagy2015economic,fosso2013implicit,beylkin2013nonlinear,maloney2002wavelets,beylkin1999fast,niklasson2002multiresolution,goedecker1999frequency,niklasson2002multiresolution,bachmayr2012adaptive,yanai2015multiresolution,EW16,EW16b,FW15,keinert2003wavelets,alpert1993class,chui1996study,johnson2001solution,alpert2002adaptive,bischoff2011low} to 
produce atom-centered adapted orthogonal bases with good completeness and locality.  
This approach requires only a modest additional
number of basis functions to yield the same accuracy as PPWs, but with a smooth,
local, and orthogonal basis.

\section{Background}
\label{back}
\subsection{The one dimensional Hamiltonian}
\label{1DHam}

Our non-relativistic many-electron Hamiltonian,
expressed in second quantized form, either in a basis set or on a grid,  is\cite{raimes1972many,helgaker2014molecular}
\begin{equation}\label{coarseHam}
\hat{\mathcal{H}}^\mathrm{MB}=\sum_{i,j,\sigma}\left(t_{ij}\hat c^\dagger_{i\sigma}\hat c_{j\sigma}+\sum_{k,\ell,\sigma'}V_{ijk\ell}\hat c^\dagger_{i\sigma}\hat c^\dagger_{j\sigma'}\hat c_{\ell\sigma'} \hat c_{k\sigma}\right),
\end{equation}
with fermionic operators $\hat c$ labeled either by site or 
basis-function $i,j,k,\ell$ and with spin $\sigma$ (or $\sigma'$).  
We define the `exact' solution as solving this Hamiltonian on a very fine
grid, which is close to the continuum limit.\cite{WSBW12,BSWB15}
For both the grid and for basis functions, we find the exact many-particle
ground state of these 1D reference systems using DMRG.
The one-electron integrals are
\begin{equation}
    t_{ij}=\int d\br\,\varphi^*_i(\br)\left(-\frac{\nabla^2}{2}+v\ext{}(\br)\right)\varphi_j(\br),
\end{equation}
where $\nabla^2=\partial_x^2$ for the 1D calculations, 
$v\ext{}(\br)$ is the external potential, discussed below.
In a basis, with functions $\varphi_i(\br)$,
the two-electron integrals are
\begin{equation}
V_{ijk\ell}=\frac12\iint d\br d\br'\varphi_i^*(\br)\varphi_j^*(\br')
v\ee(\br-\br')\varphi_k(\br')\varphi_\ell(\br).
\end{equation}
On a grid, the interaction takes a much simpler diagonal form with $i=\ell$ and $j=k$, with the integral
taking the value $v\ee(\br_i-\br_j)$.
For grid calculations, we use the ITensor library, along with matrix product operator technology.\cite{ITensor}
In the basis, we use the Block DMRG code since it is specifically
tailored to avoid stationary states that are not the ground state in a basis
set and has implemented the form of the Hamiltonian efficiently.
\cite{chan2002highly,chan2004algorithm,ghosh2008orbital,sharma2012spin,olivares2015ab}

Previously, we have explored 1D potentials which mimic as closely as possible the
behavior of real 3D systems. A particularly convenient choice matching a number of 3D features
is a single exponential function,
$v\ee(x-x')=A\exp(-\kappa|x-x'|)$ with $A=1.071$ and $\kappa=0.419$, and
$v\ext(x)=-Zv\ee(x)$, where $Z$ is the atomic number, just as in 3D.
This
particular function closely mimics the results from 
a soft-Coulomb interaction, but at a reduced
cost for grid DMRG calculations.\cite{BSWB15,ITensor} 
This potential also more closely mimics 3D since it has a mild singularity at zero distance.  In 3D, the
Coulomb interaction is divergent, but its effect is moderated, and integrals over it are finite, because 
of the very small volume associated with the $r \to 0$ region, and the associated integration factor $4\pi r^2$.
In 1D, we get qualitatively similar behavior from the  slope discontinuity in the potential at $r=0$.
A local density approximation (LDA) was also derived for
this interaction.
Our finite difference grid Hamiltonian looks like
an extended Hubbard model,\cite{WSBW12}
\begin{eqnarray}
\hat{\mathcal{H}}^\mathrm{fine}&=&\sum_i\left(-\frac1{2\a^2}\left(\hat c^\dagger_{i+1}\hat c_i-2\hat n_i+\hat c^\dagger_i\hat c_{i+1}\right)\right)\nonumber\\
&&+\sum_iv_i\hat n_i+\sum_{i,j}\Big(v\ee^{ij}\hat n_i(\hat n_j-\delta_{ij})\Big)
\label{fineHam}
\end{eqnarray}
where the superscript ``fine" indicates we will use this lattice on the finest
(original) grid of spacing $a=1/32$, $\hat n_i=\hat c_i^\dagger\hat c_i$,
external potential $v_i$, and long-ranged electron-electron interaction
$v\ee^{ij}$ on sites $i$ and $j$. 
A distance of 60 
from the outermost grid points to the first or last atom is used for all
systems that follow, allowing wavefunctions to have extended tails.

\begin{figure}
\includegraphics[width=0.9\columnwidth]{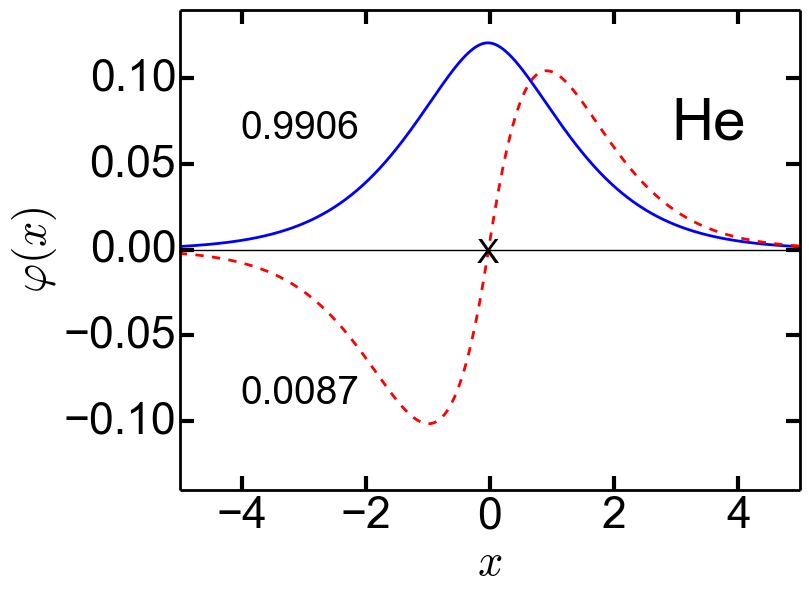}
\caption{(color online) First two natural orbitals, labelled by their occupation
numbers, of (1D) He.  An X marks the location of the nucleus.}
\label{sampleNOs}
\end{figure}
The natural orbitals are the eigenvectors of the one-particle reduced density matrix (RDM),
which is the equal-time one-particle Green's function, with matrix elements:
\begin{eqnarray}\label{denmat}
    \rho_{ij} = \langle \hat c^\dagger_i \hat c_j \rangle.
\end{eqnarray}
The eigenvalues of $\rho_{ij}$ are the occupation numbers and 
the eigenvectors are the natural orbitals, which
we order in decreasing occupation.
Fig. \ref{sampleNOs} shows the first two for 1D He, and we later show (Fig. \ref{EsampleNOs})
that, in a basis
set of these 2 orbitals alone, the expectation value of the Hamiltonian is only 1 kcal/mol
above the exact ground-state energy.
We use the term high accuracy to indicate errors of less
than 1.6 mHa, which corresponds to the 1 kcal/mol criterion commonly
called ``chemical accuracy" in quantum chemistry.

\subsection{Wavelets}
\label{s:wavelets}

Wavelets were originally introduced by Haar in 1910\cite{haar1910theorie} but they have since been modernized and expanded by several works by Gabor, \cite{gabor1946theory} Grossman and Morlet,\cite{grossmann1984decomposition} Meyer,\cite{meyer1989orthonormal} Mallat,\cite{mallat1989theory} and 
Daubechies\cite{daubechies1992ten,daubechies1988orthonormal,daubechies1993orthonormal} and many others.
These functions have become widely 
used in audio and image compression (such as jpeg and mp3 file formats).  
These were also connected to a quantum gate structure, tensor network algorithms, and compression of matrix product states.\cite{EW16,EW16b,FW15}

Consider a localized function $f(x)$ located near the origin. We can form a basis from this function by
translating it by all integer translations, i.e., $\{f(x-j)\}$ for integer $j$.  A wavelet transformation (WT) is a mapping of $f(x)$ to an new function $f'(x)$ defined by
\begin{eqnarray}
f'(x) &=&\sum_k c_k f(xd-k),
\end{eqnarray}
where $d$ is the dilation factor, which is normally taken to be 2.  The WT is defined by the coefficients $c_k$.  We will only consider compact wavelets, for which the number of nonzero $c_k$'s is finite. The scaling function of the WT, $S(x)$,  is the fixed point of this mapping. The $c_k$ are chosen cleverly to make the $S(x-j)$ to be orthogonal for different $j$, and to have a number of other desireable properties,
such as polynomial completeness up to a certain order.\cite{daubechies1992ten}
The scaling function is designed to represent smooth, low momentum parts of functions. 
The scaling function is not a wavelet, although it does form the top layer of a wavelet basis.  
A wavelet is formed from $S(x)$ using another set of coefficients $w_k$ (which are defined in 
terms of the $c_k$):
\begin{eqnarray}
W(x) &=&\sum_k w_k S(xd-k).
\end{eqnarray}
The wavelets capture higher momentum features.

\begin{figure}[htb]
\includegraphics[width=\columnwidth]{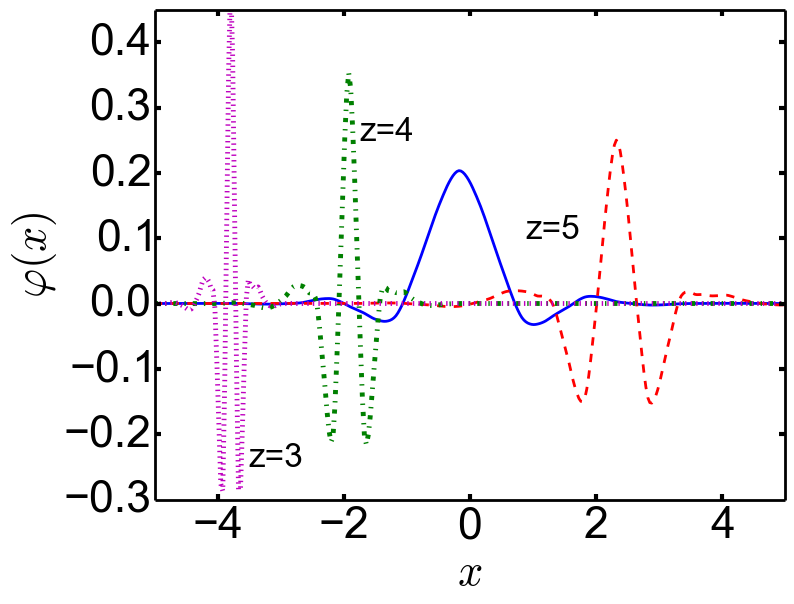}
\caption{(color online) One of the scaling functions (solid blue line) and some of the wavelets (dashed lines) of a wavelet
basis of type Coiflet-18.  These functions are based on a fine grid with spacing $1/32$, and the level parameter $z$ gives the size-scale of each function as $2^z/32$.   Both the scaling function and rightmost wavelet are at $z=5$.
}
\label{coif18}
\end{figure}

A wavelet basis consists of scalings and translations of $S(x)$ and $W(x)$, 
and it is complete and orthonormal.   
It is characterized by a coarse grid with spacing $\Delta$.  
At all integer multiples $j$ of $\Delta$, one puts a scaling function, of size $\Delta$,  
namely $S(x/\Delta-j)/\sqrt{\Delta}$.  
Then, at scales $\Delta$, $\Delta/2$, $\Delta/4$, etc., one puts down a grid of scaled wavelets, with the 
spacing and the size of the functions always equal. 
All these functions together are complete, and they are all orthogonal to each other. 
Some of the functions of a wavelet basis are shown in Fig. \ref{coif18}.

Wavelet bases are an attempt to have locality in both
space and momentum simultaneously, as much as possible, subject
to the constraint of orthgonality. The layer of scaling functions represent all momenta 
from 0 to roughly $O(1/\Delta)$; the coarsest layer of wavelets
represents momenta from roughly $O(1/\Delta)$ to $O(2/\Delta)$, etc., but with significant overlap in the momentum
coverage between different layers.

We have briefly described wavelet bases in terms of continuous functions, but they can equally be described in terms of WTs acting on an initial fine grid.  The WTs we use are based on the fine grid used by
the grid DMRG calculations, and these are what is shown in Fig. \ref{coif18}.

Many different types of wavelet transforms have been constructed. Here we choose Coiflets, derived by
Daubechies,\cite{daubechies1992ten} which are characterized by the number $\nu$ of nonzero $c_k$.  We choose relatively high $\nu$ to get good completeness and smoothness.  Wavelets can be easily extended to higher dimensions  by taking  products  such as $S(x) S(y) S(z)$,\cite{keinert2003wavelets} so the principal features of 1D carry over to 3D.\cite{bellman1957dynamic,beylkin2002numerical,rust1997using,powell2007approximate,beylkin2002numerical,reynolds2016optimization,grasedyck2013literature,anderson2010wavefunctions,bachmayr2012adaptive}

\section{Product Plane Waves}
\label{section:ppw}

In this section, we describe our new approach to 
design a specific system-dependent basis with as few functions as is practical.
We first argue that the exact natural orbitals provide a
natural least possible number, but rely on knowing the
exact solution.\cite{lowdin1956natural,lowdin1955quantum}  We then
show how to combine planewave-type basis functions (PPWs), wavelet technology, and
adaptation via approximate DFT (or other) single-particle orbitals, to create
a basis with no more than about twice this number, but still yielding high
accuracy.  A crucial feature is that we never use more than a few of each kind
of function, so that we never come close to being limited by the asymptotic
convergence properties of any one set of basis functions. Further, the initial orbitals do not need to be obtained to high accuracy.  The purpose of these orbitals is to find the important features (where the density is large) features of the system to act as a scaffold for the following calculations.  These orbitals can be obtained quickly at a low accuracy.

\subsection{Natural orbitals as a basis}
\label{NObasis}

We wish to find basis sets that, when solved exactly, give ground-state energies of high accuracy, i.e., no more than 1 kcal/mol above the exact, complete basis limit. 
We wish to find basis sets that converge to this accuracy with as few functions as
possible, but also without needing to know the exact solution to determine them.
With the fine grid DMRG wavefunction, we
can calculate $\rho_{ij}$ exactly and find the exact natural orbitals. 
Since our DMRG solutions do not break spin symmetry
if the number of electrons is even, the up and down RDMs are identical.
(For odd electron numbers, we average the up- and down-RDMS and use that to define our natural
orbitals.)

\begin{figure}[htb]
\includegraphics[width=\columnwidth]{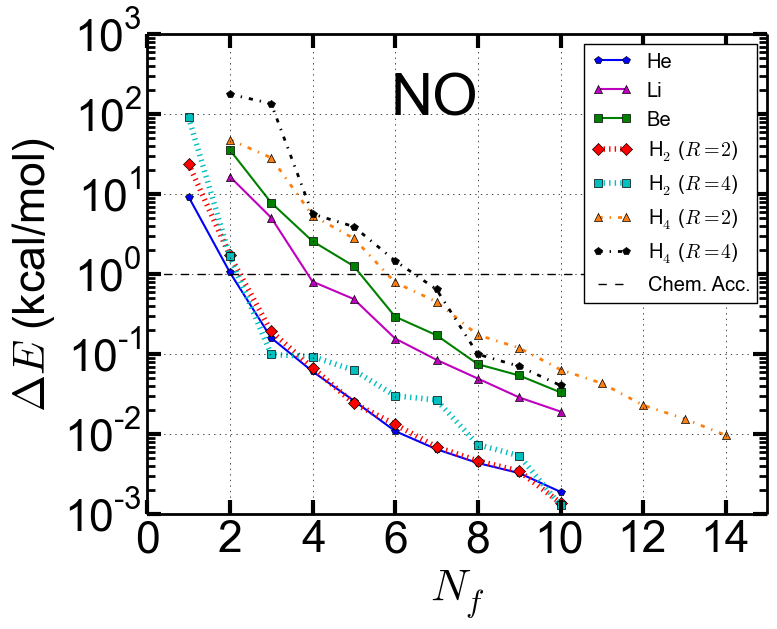}
\caption{(color online) Energy errors for 1D
He, Li, Be, H$_2$, and H$_4$ when evaluated in a basis of
$N_f$ exact NOs of greatest occupancy.  
}
\label{EsampleNOs}
\end{figure}
The first two natural orbitals for a 1D helium atom were shown in Fig.~\ref{sampleNOs}.
The natural orbitals yield the smallest number of basis
functions that can be expected to yield high accuracy, i.e., when ordered
by occupancy, the least number $M_{no}$ which, when used as a basis, yields an error
below high accuracy.
Fig.~\ref{EsampleNOs} shows the energy error for a variety of systems, when the basis is
chosen as a finite number of the most occupied exact natural orbitals.
We see that $M_{no}=2$ for 1D He, but is 3 for 1D H$_2$ either close to equilibrium ($R=2$)
or stretched ($R=4$).  For 1D Li, $M_{no}=4$, while 1D Be has $M_{no}=6$. 
Unstretched 1D H$_4$ also has $M_{no}=4$, but stretched 1D H$_4$ requires $M_{no}=7$.  
Thus $M_{no}$ increases with the number
of electrons, and also (slightly) with the number of centers.

\begin{figure}[htb]
\includegraphics[width=0.9\columnwidth]{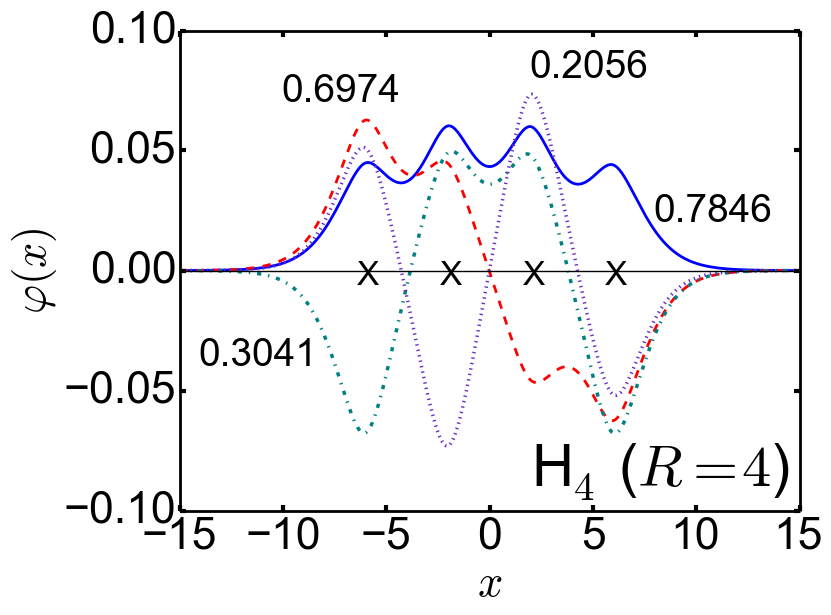}
\caption{(color online) Same as Fig. \ref{sampleNOs} but for 1D
H$_4$ at $R=4$.  X's mark the locations of the nuclei.
}
\label{H4NOs}
\end{figure}
Fig. \ref{H4NOs} shows the first four natural orbitals for an 1D H$_4$ chain,
which is stretched.  Clearly, the orbitals delocalize over the entire chain.
We also see from Fig. \ref{EsampleNOs} that even in this basis, there remains about
8 kcal/mol error, and 3 more orbitals are needed to reach high accuracy.

\subsection{Constructing the basis}
\label{PPW}

Given the orbitals from a HF or DFT calculation, perhaps the simplest conceivable basis would be 
the occupied HF or DFT orbitals, since this
allows the reproduction of the single determinant.
One well-known approach for enlarging this basis to allow for correlation is to use additional eigenstates
of the Fock matrix, selected by an energy cutoff.\cite{glover2010first,anderson2010wavefunctions,flad2002wavelet,anderson2010wavefunctions}  It is clear, however, that this eventually 
becomes inappropriate.  For a more complete basis, one needs functions with positive energy, 
but there are an infinite number of functions at zero energy far from the molecule.  To remedy this, we could put
a box around the molecule and include only functions within that box.
However, this can be very wasteful, since the
box needs to include extended tail regions, where additional basis functions are not very useful.
Instead of using energies, we adopt a quite different approach, motivated by the construction of variational
wavefunctions---in particular, Jastrow functions.

Single-particle determinantal states $\varphi$ from DFT or HF are rough approximations
to the many-particle wavefunction, but 
can be improved substantially by multiplication by a 
Jastrow factor, $\cal J$, 
which provides explicit correlation.  Modifying a determinantal wavefunction with a Jastrow factor
is often the first step in designing a variational wavefunction for quantum Monte Carlo calculations\cite{nightingale1998quantum}
The Jastrow factor acts as a multiplicative factor for the
wavefunction and simple form for $\cal J$ is\cite{umrigar1988optimized}
\begin{equation}
{\cal J}(\br_1, \br_2, \ldots) = \prod_{i<j} J_2(\br_i-\br_j).
\end{equation}
The $J_2$ term is near 1 if $\br_i$ and $\br_j$ are far away, and becomes less than one
as $\br_i$ and $\br_j$ come together, building in the electron-electron cusp.  We now ask
the question:  what would be a good {\it single-particle basis} to represent $\cal J$ or $J_2$?  

The fact
that $J_2$ is a function of the difference of two position vectors means that there is no benefit
to increasing resolution in one region relative to another, at least for fitting $J_2$.  One does expect,
however, that longer wavelength functions are more important than short wavelength functions. This
suggests that a plane wave basis, restricted to the general vicinity of the molecule, with a momentum
cutoff which is not too high, is a reasonable approximate basis for a Jastrow function.

Since the Jastrow function in a variational wavefunction multiples the determinant
of occupied DFT orbitals, 
this suggests a very simple ansatz for a basis for correlated calculations:  the product of occupied
orbitals and low momentum cutoff plane waves, which we call a product plane wave (PPW).
To be more specific: let $\{ b_k(\br)\}$ be a set
of plane waves with a low momentum cutoff, and let $\{ \varphi_j(\br) \}$ be the occupied orbitals
from a DFT/HF calculation.  Then our product plane-wave (PPW) basis is $\{ \varphi_j(\br) b_k(\br) \}$.
The momentum cutoff in $\{ b_k(\br)\}$ corresponds to some minimal resolution.  Linear combinations
of the $b_k(\br)$ can represent a correlation hole at any position within the system, while high momentum behavior
near the nuclei is captured by the $\{ \varphi_j(\br) \}$.  $b_{k=0}=1$,
so that the $\{ \varphi_j(\br) \}$ themselves are part of the basis.

\begin{figure}
\hspace{0.15cm}\includegraphics[width=0.97\columnwidth]{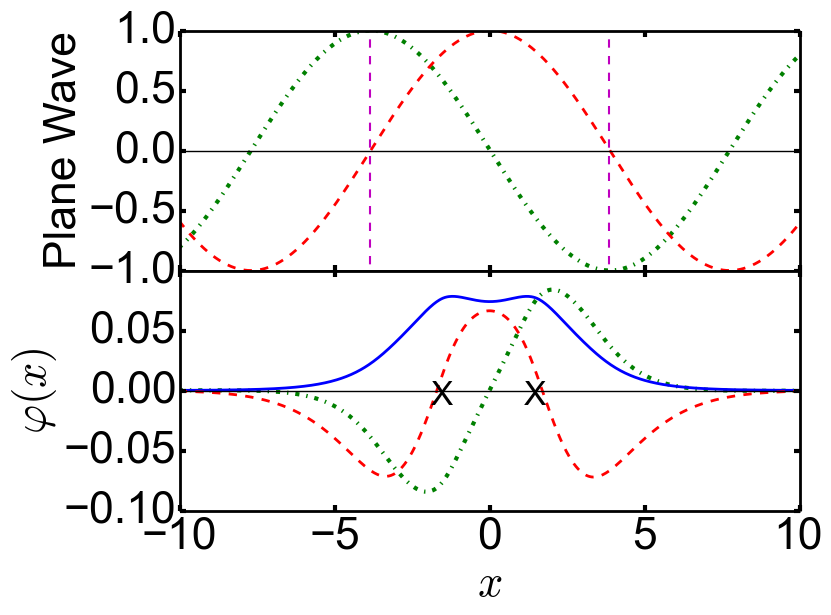}
\caption{(color online) Product plane wave (PPW) functions
for a 1D H$_2$ at $R=3$.  Here the box size is $L=7.72$,
and marked by pink vertical lines. The upper figure shows
the windowing functions $\cos(k_1x)$ and $\sin(k_1 x)$ and the lower figure shows
the first three PPWs (the first is just the LDA orbital). 
}
\label{PPWBresults}
\end{figure}
In generating a PPW basis, several choices must be made.  First, we want to put the molecule in a ``box'' 
that defines the sequence of momenta in the plane waves.  Since the detailed correlations we want from
the plane waves are weak in the tails, and since the box size is only used to define momenta, 
we do not include long density tails.  We simply choose a small density cutoff, $\rho_m$, to
define the edge of our box, from our DFT (or HF) calculation.
Here $\rho_m=10^{-3}$ throughout, but we expect our qualitative
results to be very insensitive to this choice.  For neutral atoms, the corresponding box sizes are
$4.90, 5.34, 8.40,$ and $8.71$ for $Z=1$ to $4$.
A simple example of a product plane wave basis is illustrated in Fig. \ref{PPWBresults}.
The first two functions resemble the natural orbitals of 1D He in 
Fig. \ref{sampleNOs} and the natural orbitals here. 
This resemblence between PPWs and NOs tends to continue for 
higher functions, although the precise order of the functions can vary.

\begin{figure}
\begin{center}
\includegraphics[width=0.89\columnwidth]{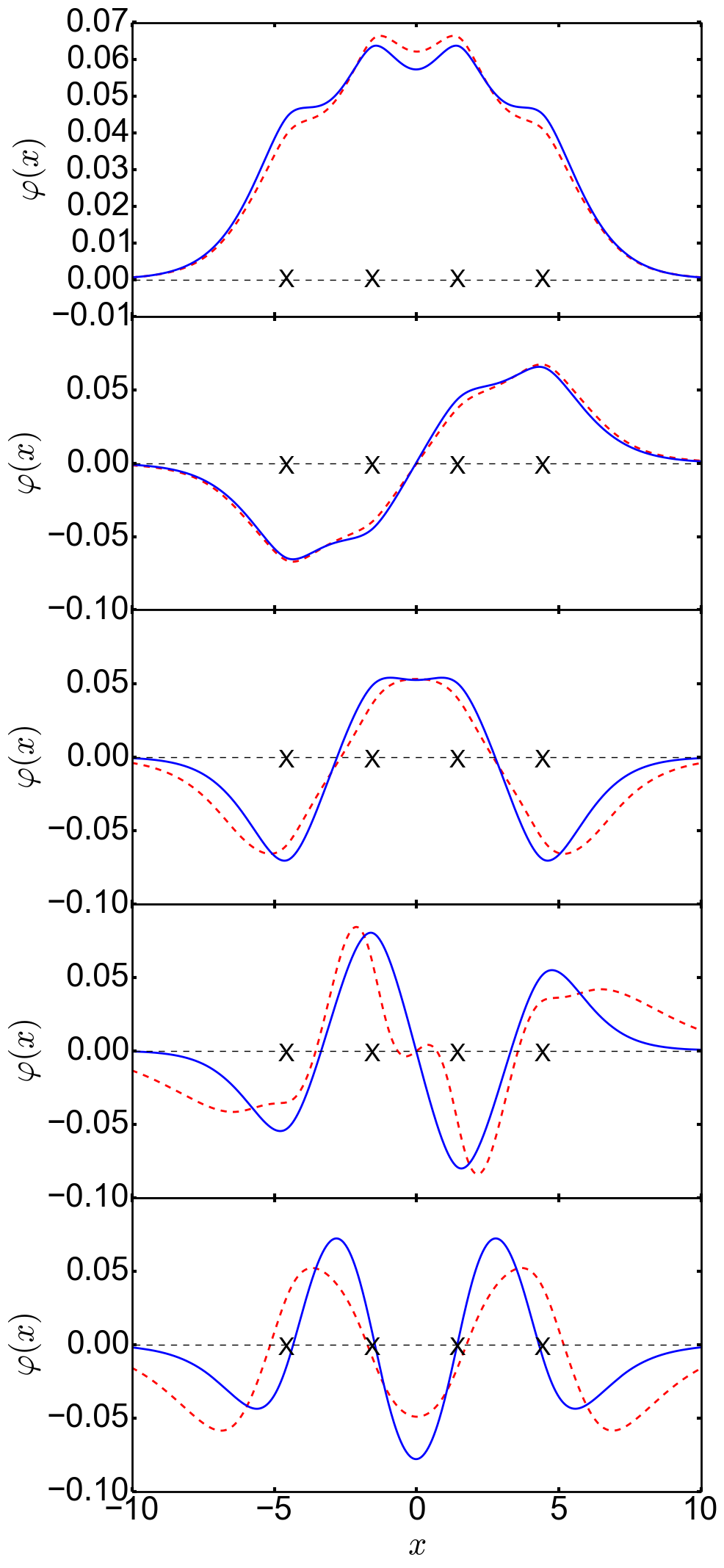}
\end{center}
\caption{(color online) The first five PPWs (red dashed) after orthogonalization compared to the exact
natural orbitals (blue) for 1D H$_4$ with
$R=3$. Here $L=13.8$.  These functions are similar to those found
in Ref.~\onlinecite{LBWB16} from density-partitioning. }
\label{fctfig}
\end{figure}
Let $\jindex$ be the number of
occupied orbitals in a DFT or other approximate calculation.  
Let $L$ be the width of the box defined by the cutoff $\rho_m$. 
Then choose an integer $J\ge 0$ to create $2J+1$ functions, the identity and
$\cos(k_n x)$ and $\sin(k_n x)$,
where $k_n=n\pi/L$, $n=1,\dots,J$,
and multiply each by the occupied DFT orbitals, creating $(2J+1)\jindex$ primitive PPWs.
Next, we exactly orthogonalize these orbitals via the Gram-Schmidt process, 
in the order of $k$-values, starting with the
identity.
The results for 1D H$_4$ are shown in Fig. \ref{fctfig} and compared to the exact natural orbitals.
These orthogonalized PPWs are remarkably close to the exact natural orbitals, 
especially for those orbitals that are occupied in the DFT 
calculation, but also even for those that are not.  (The additional wiggle in the 4th PPW 
is due to the orthogonalization procedure).

\begin{figure}
\includegraphics[width=\columnwidth]{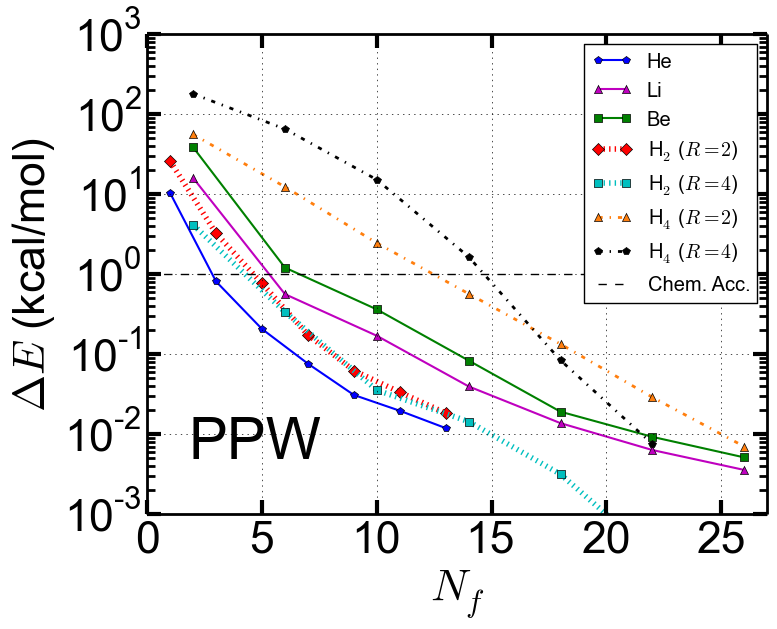}
\caption{(color online) Finite-basis error of PPWs, to be contrasted with
Fig. \ref{EsampleNOs}, which has exact NO's.
\label{EsamplePPWB}}
\end{figure}
Finally, in Fig. \ref{EsamplePPWB}, we show the energies for our systems as a function of the number of
(orthogonalized) PPWs.
For 1D He and 1D H$_2$, $\jindex=1$, so increasing $J$ by 1 yields two more PPWs (the sine and the cosine);
for the rest, $\jindex=2$, and 4 PPWs are added each time.
A quick glance shows a remarkable similarity to the ordered natural-orbital energy errors of Fig. \ref{EsampleNOs}.
The PPW functions yield high accuracy with a few more functions than $M_{no}$,
showing that they do not just look similar to the NO's, they are similar in an energetically meaningful
sense. We denote $M_{PPW}$ as the least number needed to reach high accuracy.
A more careful inspection shows that they are not quite as accurate, even for 1D He, and that the difference
grows with the number of electrons and the number of atoms.  It is most noticeable for stretched 1D H$_4$, where
$M_{PPW}=18$, whereas $M_{no}=7$.  But this is still a remarkably small number
for a strongly correlated system.

\section{Wavelet localization}
\label{local}

So far, we have accomplished our goals of a basis function set with a low number of orbitals.  Our PPWs yield high accuracy with about $2M_{no}$ basis functions.  But, to be
efficient, tensor network
methods such as DMRG require the low entanglement that
comes from localized basis sets. Other methods may also
benefit from localized basis functions, which make Hamiltonians sparse.  Now we study
cases with more than one atom, showing how we can use wavelet technology to break down a PPW
into localized, smooth orthogonalized basis functions, centered around each atom,
without too large an increase in the number of functions.

Traditional methods for localization rely on orthogonal transformations within the set of basis functions one
already has.  Not enlarging the set of functions puts a
strong limit on how localized the functions can be made.
However, if one enlarges the space without limit, one can make the basis as local as one wishes.  
One can think  of ``chopping up" each delocalized basis function 
(which we can picture as a molecular orbital):  
partition all of space into a chosen number of
disjoint regions, or {\it cells}.\cite{frediani2015real,losilla2012divide,losilla2013numerical,sherrill2010frontiers,losilla2013numerical,almlof1991atomic,widmark1990density}  
For example, one can make the number of cells the same as the number of atoms, and define
each cell by associating each point in space with the closest nucleus. 
Form a basis by projecting each delocalized basis function into each cell, i.e. multiplying it by a function
which is unity for points in the cell and zero outside, and repeating for all delocalized functions.  
Linear combinations of 
chopped up functions would allow one to reproduce any of the original delocalized functions, but
this would make a terrible basis, for two reasons:
1) discontinuous basis functions have infinite kinetic energy, and
2) the number of localized functions
scales as the square of the number of atoms.

Using wavelets, we can retain this idea of
``chopping up" basis functions into different regions, but fix these two
problems.  As discussed in \ref{s:wavelets},
we define a complete wavelet basis consisting of a grid of scaling
functions
with lattice spacing $\Delta$  (say with $\Delta \sim 1$ Bohr), 
and an infinite sequence of wavelets at scales
$\Delta$,  $\Delta/2$,  $\Delta/4$, etc, as shown in Fig. \ref{coif18}.  
We will refer to any of these functions,
either a scaling function or a wavelet of any scale, as a WF (wavelet-function). 

Now to chop up a delocalized basis: expand all delocalized functions in terms
of the WFs.  Many WFs will not have significant overlap with any functions, and
can be dropped.  This procedure thus produces a localized but
smooth basis encompassing the original functions,
assuming one has chosen smooth wavelets.  However, the number of functions tends to be rather high, so we
use this only as a starting point.

\begin{figure}
\includegraphics[width=0.9\columnwidth]{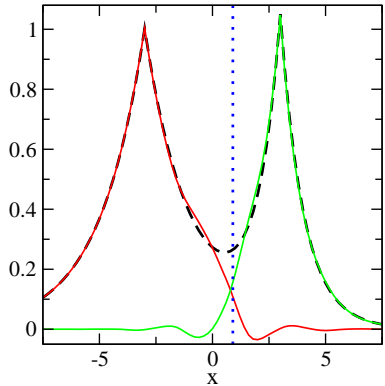}
\caption{(color online) The function $\exp(-0.5*|x+3|) + \exp(-|x-3|)$ (black dashed) is divided into two 
orthogonal pieces (red and green solid lines) using wavelets. 
The wavelet basis used was based on Coiflet-24 with $\Delta=1$, and the dividing line separating
the two cells (dotted line) was $x=0.9$. The small oscillating tails make the two function pieces orthogonal.
The two singularities each only appear in one piece, because the high momentum wavelets representing the
singularities are more and more localized the higher the momentum. \label{chopping}
}
\label{wavelets}
\end{figure}

Again we partition all of space into cells, associated with atoms.
Associate each WF to a cell.
A natural way to do this is to define a center of mass
for each function, and then the WF goes in the cell that
contains its center of mass.
Now we can project each delocalized function into each cell, simply by expanding the function in terms of
the WFs belonging to the cell.  This cuts the delocalized function into pieces which are all orthogonal.  
An example of this procedure is shown in Fig. \ref{chopping}.

If we repeat this with additional delocalized functions, the pieces in different cells will be 
orthogonal, even if they
came from different delocalized functions, since the WFs of different cells are orthogonal.  
Within a single cell,
the pieces will not be orthogonal, and may have substantial overlap.
The final step is to recombine all the pieces in a particular cell into a
reduced set of orthogonal functions 
for that cell, and repeat for all cells.  
Note that while the original delocalized functions may
be normalized, the pieces come from a projection and will not be, and some pieces may have 
very small normalization. It is important to leave the pieces unnormalized.
For each cell, we wish to find the minimal set of basis functions that can
represent all the pieces to within a specified accuracy. 
This is a well known linear algebra problem with a simple solution. Let $f^i_j$ be the piece of delocalized function
$i$, expanded in terms of the WFs $j$ belonging to a cell $c$.  Form a cell covariance matrix
$\rho^c$ as
\begin{eqnarray}\label{corrmatrix}
    \rho^c_{jj'} \equiv \sum_i f^i_j f^i_{j'}
\end{eqnarray}
Then the reduced basis we seek is the set of eigenvectors of $\rho^c$ (which is positive semi-definite) 
with eigenvalues above a specified cutoff, $\eta$.  This cutoff is roughly the mean-square error in representing all the
different pieces.  This is often called a principal component analysis.
\cite{wold1987principal,abdi2010principal,vu2015understanding,li2016understanding,LBWB16}  Here we call the entire process wavelet localization (WL) and
the resulting basis functions wavelet-localized orbitals (WLOs).  Although the WLO procedure
could be applied to other delocalized bases, here we will only consider its application to
PPWs.

\begin{figure}
\includegraphics[width=\columnwidth]{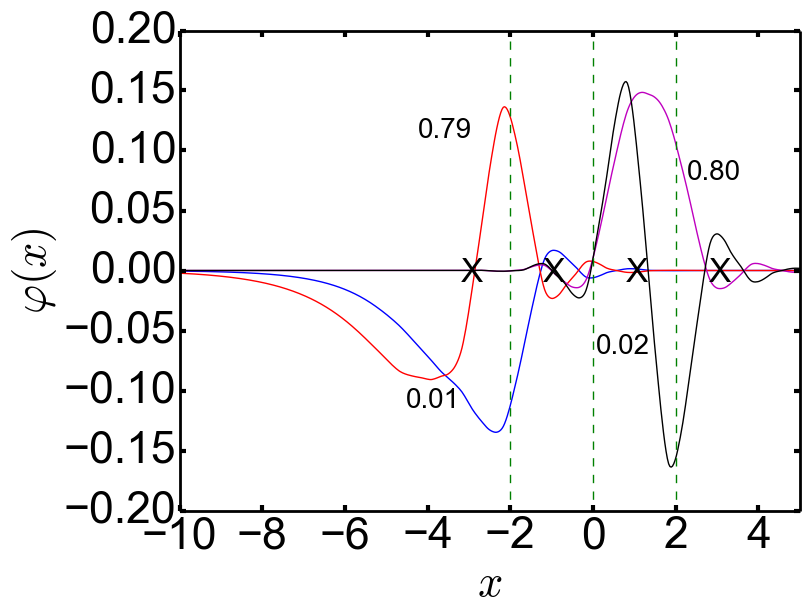}
\caption{Some of the WLOs for each cell of a 1D H$_4$
chain.  Shown are the first two in both the first atom's cell (far left) and
third atom's cell.  Green vertical lines are drawn midway between each atom
and weights of each function are labeled near each curve.  The calculation this
was taken from was $b=0$, $\Delta=1.0$, $N_J=0$.}
\label{separatedfunctions}
\end{figure}
Fig. \ref{separatedfunctions} shows the results of wavelet localization for 1D H$_4$, 
with a spacing $R=2$, discussed more in Sec. \ref{performance}.
For simplicity, the figure shows only the two leading
eigenfunctions and their eigenvalues for only cells 1 and 3. 
The dashed lines show the dividing lines between
the different boxes; the nuclei are  at $x=-3$, $-1$, $1$, and $3$.
The functions are all orthogonal, with oscillations in the tails of each function to ensure orthogonality
between boxes.

The parameter $\Delta$, the spacing of the scaling functions, is crucial, as it sets the size of the region in
which functions on adjacent boxes overlap. In the limit $\Delta \to 0$, this
chopping up procedure reduces to
the naive discontinuous procedure mentioned at the beginning of this section.
The procedure also becomes poorly behaved if $\Delta$ is larger than the interatomic spacing.  Roughly, one
should set $\Delta$ to a modest fraction of the interatomic spacing, but later on we show results as a 
function of $\Delta$ to determine optimal values.

Lastly, we note that, for multi-center stretched systems, 
if $R>L_a$, the box for an atom, then we use $L_a$ instead of $L$ for that cell.
This can greatly increase the number of functions to $N_a\times N_{fa}$,
where $N_{fa}$ is the number needed to reach high accuracy
for the isolated atom, but unneeded functions will be discarded by our wavelet localization.

\subsection{Performance of WLO bases}
\label{performance}

In this section, we wish to check that our WLOs work well for some correlated quantum calculations,
and find out how many WLOs are needed for a given task.
Our procedure requires, at most, $N_{occ} \times (2J+1) \times N_{cell}$ functions.
Thus, for a H$_4$ chain that
is unstretched (no spin-symmetry breaking), $N_{occ}=2$, we will usually choose $J=1$, and have 4 cells.
A PPW calculation has 6 functions, and up to 24 (6 per cell) when fragmented.  However, in practice,
up to half those functions can be eliminated by the cutoff of our covariance matrix.
This removal of irrelevant functions becomes increasingly important as the number of atoms grows.

\begin{figure}
\includegraphics[width=\columnwidth]{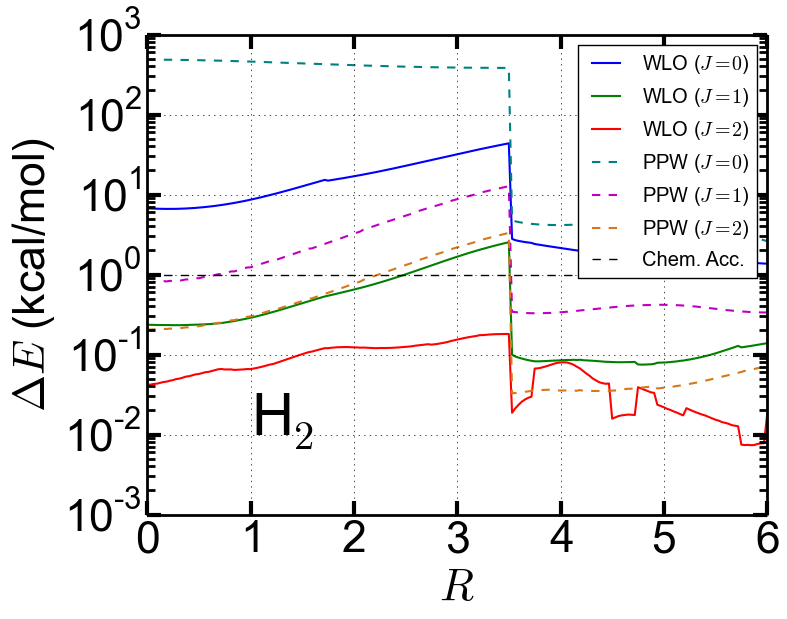}
\caption{Error as a function of bond length $R$ for 1D H$_2$ using both pure PPWs
and WLOs, for various values of $J$. 
The sudden shift is at the Coulson-Fischer point of the LDA calculation,
beyond which a broken spin-symmetry solution, with twice as many orbitals,
has the lowest energy.
}
\label{dEH2}
\end{figure}
The prototype calculation is the dissociation of molecular hydrogen.  All single-determinant methods fail
as bonds are stretched and electrons localize on distinct sites.  Molecular hydrogen dissociates into an
open-shell biradical (two 1D H atoms).  The molecular energy as a function of separation is given in 
Fig 6 of Ref.~\onlinecite{BSWB15}.
That figure also shows the failure of LDA, with a Coulson-Fischer point\cite{coulson1949xxxiv} $R_{CF}=3.53$, 
where the unrestricted broken
symmetry solution becomes lower in energy than the spin-singlet within LDA.  
In Fig \ref{dEH2}, we show the error in the energy curve, using pure PPWs, and also separating into separate
cells, using $\Delta=1$ and $\eta=10^{-4}$. 

Beginning with the PPWs (dashed lines), we see that increasing $J$ improves accuracy systematically,
as expected.  Moreover, for a given $J=1$ or higher, we see that the error increases systematically as the bond
is stretched until $R_{CF}$ is reached.   This is because the LDA orbital is
becoming less and less close to the exact natural orbital as the bond is stretched.  Beyond this point,
there is a great decrease in error, as the the number of LDA 
orbitals doubles (due to spin-symmetry breaking).
Even the largest PPW basis shown here ($J=2$) does not achieve high accuracy close to the CF point.
But our WLOs {\em do} reach high accuracy everywhere for $J=2$, and almost
everywhere with $J=1$, using $3\times 2  =6$ functions for $R < R_{CF}$, and double that beyond.  
(The wavelet localization does not throw out any WLOs here.)
Thus our basis set works, even through the CF point.
Of course, in practice, quantum chemists want forces, and some smoothing procedure would be adopted to
avoid the kink at the CF point.

The strong changes with $R$ in the error in the
red curve past the CF point can be attributed to
the grouping of the scaling and wavelet functions.  
As the bond is stretched, because the functions are fixed
in real space, some of the functions are assigned to the left cell,
and others to the right.  This assignment can change suddenly,
causing a drop in the eigenvalue weights in the covariance matrix of one
of the cells and decreasing the number of functions.
Note that this effect occurs only for errors far below the high
accuracy threshold.

\begin{figure}
\includegraphics[width=\columnwidth]{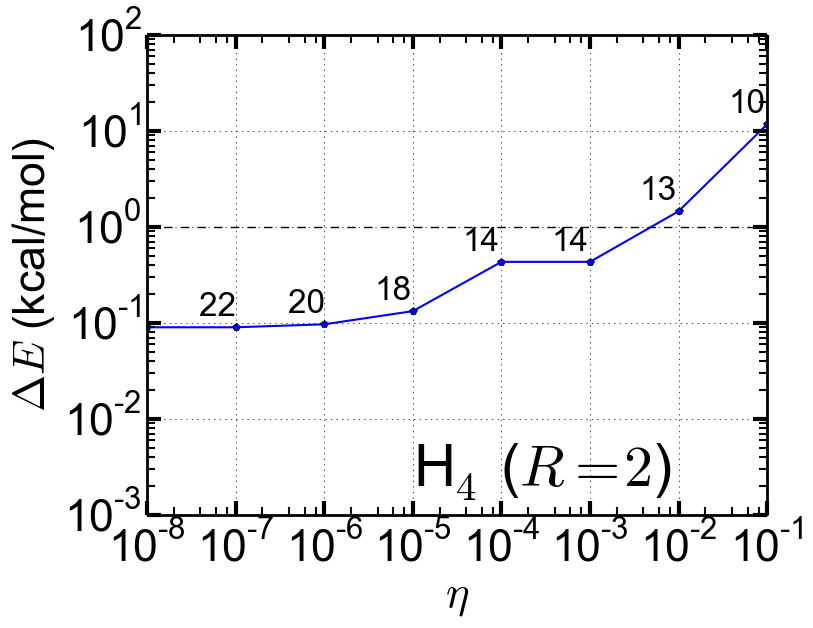}
\caption{Finite-basis energy error as a function of covariance cutoff $\eta$
for 1D H$_2$ at $R=2$ with $J=\Delta=1$. Without cutoff, there are 24
functions in the basis.  The integer near each
point is the number of functions
in the basis.}
\label{cutoffenergies}
\end{figure}
Next we consider performance for longer chains of 1D H atoms.
Now the covariance cutoff becomes important for curtailing the total
number of functions.
Figure~\ref{cutoffenergies} illustrates the effect of the covariance
cutoff for H$_4$ near equilibrium.   The higher the value of $\eta$, the more functions are thrown
away, but the greater the error is.   If $\eta$ is set too small, then
no functions are removed, not even those that have essentially no effect
on the energy.  The figure shows that the full basis has an error of about
0.1 kcal/mol.  But 
high accuracy is achieved with $\eta=10^{-3}$ and only 14 functions.
This is to
be contrasted with $M_{no}=6$ from Fig \ref{EsampleNOs} and 
$M_{PPW}=14$ from Fig.~\ref{EsamplePPWB}.
In this case (near equilibrium), the WLOs form a near-complete
localized orthogonal basis with no more functions than PPW, and
with lower error.
Note that setting $\eta=10^{-4}$ does not add in any more functions.

\begin{table}
\begin{tabular}{|c||cc|cc|cc|cc|cc|cc|cc|cc|cc|cc|cc|cc|cc|cc|cc|cc|cc|cc|cc|cc|cc|cc|}
\hline
$R$& \multicolumn{2}{c|}{$2$} & \multicolumn{2}{c|}{$3$} & \multicolumn{2}{c|}{$4$} & \multicolumn{2}{c|}{$5$}
 & \multicolumn{2}{c|}{$6$} \\
\hline
$\Delta$ & $N_f$ & $\Delta E$ & $N_f$ & $\Delta E$ & $N_f$ & $\Delta E$ & $N_f$ & $\Delta E$ & $N_f$ & $\Delta E$\\
\hline
0.5 & 16 & 0.24 & 16 & 0.33 & 26 & 0.11 & 24 & 0.09 & 23 & 0.21 \\\hline
1.0 & 14 & 0.43 & 16 & 0.26 & 24 & 0.15 & 22 & 0.11 & 22 & 0.16 \\\hline
2.0 & 16 & 0.37 & 15 & 1.50 & 28 & 0.04 & 25 & 0.08 & 24 & 0.15 \\\hline
4.0 & 18 & 0.34 & 18 & 0.52 & 29 & 0.04 & 25 & 0.10 & 25 & 0.17 \\\hline
\end{tabular}
\caption{WLO ($J=1$) errors for 1D H$_4$ as a function of separation, for various
values for $\Delta$.  Chopping the PPWs yields up to 48 functions, but
setting $\eta=10^{-4}$ as the covariance cutoff yields the number of
functions and accuracy shown.  The units provided are in kcal/mol.}
\label{H4table}
\end{table}
To see the effect as a function of bond length, in Table \ref{H4table},
we give energy errors and numbers of basis functions for various values
of $R$ and several values of $\Delta$, for a $J=1$ calculation with $\eta=10^{-4}$.
(In all cases, $J=0$ was found to yield errors higher than 1 kcal/mol.)
We see that the least number of functions needed occurs for $\Delta=1$, especially
as the chain is stretched.

\begin{table}[htb]
\begin{tabular}{|c||cc|cc|cc|cc|cc|cc|cc|cc|cc|cc|cc|cc|cc|cc|cc|cc|cc|cc|cc|cc|cc|cc|}
\hline
$R$ &  \multicolumn{2}{c|}{$1$}  &  \multicolumn{2}{c|}{$2$} &  
\multicolumn{2}{c|}{$3$} &   \multicolumn{2}{c|}{$4$} \\ 
\hline
$\eta$ & $N_f$ & $\Delta E$ & $N_f$ & $\Delta E$ & $N_f$ & $\Delta E$ & $N_f$ & $\Delta E$\\ 
\hline
$10^{-4}$ & 42 & 0.47 & 51 & 0.10 & 50 & 0.63 & 49 & 0.25  \\\hline
$10^{-3}$ & 42 & 0.47 & 43 & 1.29 & 49 & 1.08 & 40 & 0.83  \\\hline
\end{tabular}
\caption{Same as Table \ref{H4table}, but for 1D H$_{10}$, with $J=\Delta=1$,
and two different covariance cutoffs.}
\label{H10table}
\end{table}
Finally, we have run examples of 10-atom chains. We achieve high accuracy for
$J=1$, $\Delta=1$ throughout the range of $R$ shown in the table, with about 5 functions
per site when $\eta=10^{-4}$.  This may seem like a large number of functions, but
keep in mind that, as $R$ increases, this is a strongly correlated system tending toward
its thermodynamic limit.  Moreover, we have required our total energy to be accurate
to 1 kcal/mol all along the curve, not just the energy per atom.  
One would also expect most energy differences to
converge more rapidly than the total energy.
Table \ref{H10table} also illustrates the benefits of the covariance cutoff.  By setting
its value to $10^{-3}$, we significantly reduce the number of functions as $R$
increases, but in the middle, our error is slightly greater than 1 kcal/mol.
For many practical purposes, this should be sufficient, but the larger lesson is that,
for any desired application, there is a controllable trade-off between accuracy and
number of functions.

\begin{figure}[htb]
\begin{center}
\includegraphics[width=0.9\columnwidth]{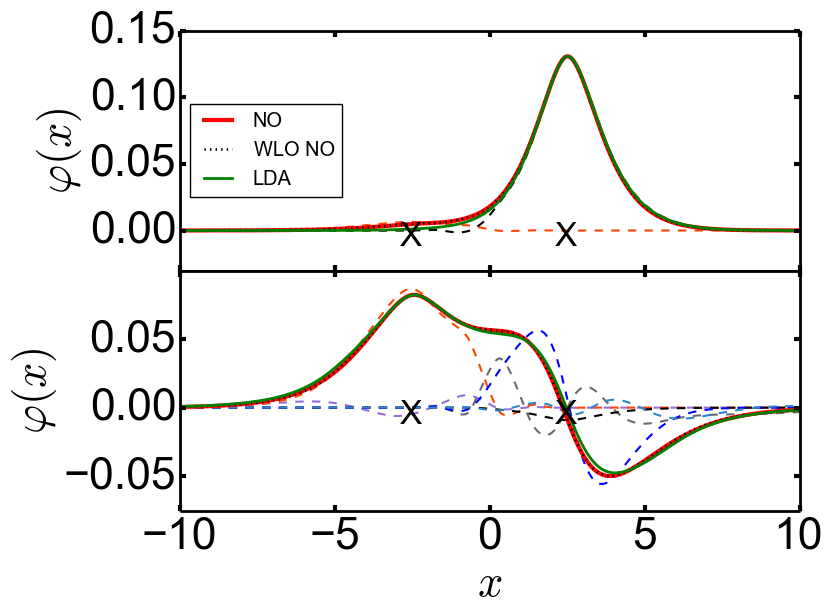}
\end{center}
\caption{The first two natural orbitals for stretched 1D LiH (X's denote nuclear centers with Li on right).
The exact NO's are marked in red, and are indistinguishable from the WLO NO's ($\Delta=J=1$, 
$\eta=10^{-4}$), black dotted line,
but slightly different from the occupied LDA orbitals (green).  Also shown are
WLOs with weights above $10^{-4}$ (dashed lines). 
The WLO basis has 11 functions, and an
error of 1.04 kcal/mol.
}
\label{LiHNO}
\end{figure}
We end with a heteronuclear diatomic, 1D LiH, to show that our method still works in the
absence of left-right symmetry.  Fig. \ref{LiHNO} was calculated with $J=\Delta=1$ and $\eta=10^{-4}$.   The LDA orbitals remain an excellent starting point for approximating the
NO's, and the NO's in the WLO basis are identical (on this scale) to the exact NO's.
The energy error is only 1.04, using 11 basis functions.

\section{Discussion and conclusions}
\label{conclusions}

We have presented algorithms to generate a basis set that is
adapted to a specific molecular system and designed to be used in correlated
calculations.  The basis begins with an inexpensive DFT or HF calculations, and
the generation of additional functions from the occupied orbitals to allow
correlation is even less expensive.
A product plane
wave (PPW) ansatz adds additional functions using a product of low momentum plane waves times each
occupied orbital. In our 1D test systems, this ansatz produces
results within high accuracy using
about twice as many functions as in an ideal natural orbital basis.
Then, to generate basis functions localized near each
atom, we introduced a wavelet localization procedure.
Compared to standard localization methods, which involve an orthogonal transformation 
of the existing functions without expanding the basis, wavelet localization
produces stronger localization with much smaller orthogonalizing tails, at the expense of
adding basis functions. This procedure is particularly useful for DMRG calculations,
where locality in the basis is an important criteria.
It may also improve scaling on large systems in other correlation approaches.
Our method, as presented here, should allow much larger systems to be treated
than previously possible in our 1D mimic of realistic electronic structure
(such as the 100-atom chains of Ref.~\onlinecite{SWWB12}).

Our procedure has only been given and tested upon a 1D mimic of the 3D world.
A naive generalization of PPW to arbitrary 3D problems would involve many more 
plane waves, roughly the cube of the number in 1D.  For a fixed momentum cutoff the number
of plane waves also grows with the length of the system, even in 1D. This would appear to generate
far too many functions to be practical, but the wavelet localization would counteract this
effect.  We can think about how this works by considering one particular cell, centered on
an atom.  The PPW basis generates occupied orbitals times plane waves with a low momentum cutoff.  The
number of functions needed to span this set in one cell should not be too large, since the only high frequencies
present are from the cusps of the occupied orbitals at the nuclei, which in a Gaussian basis can be represented by a small number of basis functions. Otherwise, there are only a limited number of low frequency
modes in a single atom cell.  This means that there must be significant redundancy in the PPW functions,
particularly for many electrons.  The principle component analysis of the wavelet localization would remove
this redundancy. This makes it clear that except for very small molecules, one should not apply PPW on its own, but in conjunction with wavelet localization.  Nevertheless, there are likely significant challenges in going to 3D which we must leave
for future work. In 1D, our bases give high accuracy with only about twice as many functions as in an equivalent natural orbital basis.
It  seems reasonable that a variation of our 1D approach can be found for 3D which is similarly less efficient than a natural orbital basis by only a modest factor.

In the case of a He atom, this means roughly that 3D He would need about the cube of the number of functions as 1D He.  This argument would apply to any basis, including natural orbitals.  Indeed, one finds one needs about 15 NOs for chemical accuracy in 3D He,\cite{ahlrichs1966solution} versus 2 or 3 for 1D He.  Our PPW basis does not try to beat the NOs, which is not possible; rather, it tries to duplicate their completeness but based on a cheap calculation. In 1D, we obtain the same accuracy as with an NO basis if we use about twice as many functions.  In 3D, we hope to do similarly--but this has not been tested.

One improvement to our PPW approach which we have not explored here is to give more weight to
the occupied orbitals than to the additional functions coming from the plane waves with nonzero
momentum. This would be fairly simple to implement in our wavelet localization, by multiplying
the $J>0$ functions by a weighting factor less than 1.  One would expect this natural modification
to further reduce the number of functions needed for high accuracy. 
We also note that our procedure could also be applied without chopping, but still
removing irrelevant basis
functions, by constructing the orthonormal basis from the PPWs
\begin{eqnarray}\label{onecell}
g^i_j = \sum_k O^{-1/2}_{jk} f^i_k ,
\end{eqnarray}
where $O$ is the overlap matrix of the $f^i$.
Now $\rho^c = O$, so the principle component analysis 
consists of forming a basis of the eigenvectors of the overlap matrix with the 
largest eigenvectors, up to cutoff $\eta$.  This procedure 
reduces basis-set linear dependence; here it might reduce the PPW basis
size significantly without much loss of accuracy.

A number of existing approaches also utilize or are based on approximate natural orbitals.
For example, some Gaussian basis sets attempt to reproduce properties of atomic natural orbitals.\cite{neese2010revisiting} 
A key difference with our approach is that we start from the beginning
with orbitals adapted to the specific molecule under consideration, based on a DFT or HF calculation.
It would be interesting to compare the number of functions needed to reach chemical accuracy in 3D
between our PPW approach and standard Gaussian basis sets.  (We do not have these Gaussian
basis sets for our 1D test systems.)

Another common approach  is to find approximate natural orbitals from a low-order correlation
calculation, such as second order perturbation theory, e.g. MP2.\cite{gruneis2011natural}
Our PPW method is simpler and faster, and it would be
interesting to compare the accuracy of these two approaches.  One might also combine them: in cases
where the perturbation calculation was expensive to do in a large basis, one might first get a PPW
basis, which would be much smaller than an unadapted basis, and then refine it further 
by getting approximate natural orbitals with a perturbation theory approach.

The localization using wavelets could be applied in a broader context than we have used here, such as to
standard Gaussian bases or to approximate natural orbitals coming from a low order correlation method.
This could potentially improve the performance of DMRG or other tensor network methods.  By improving
the sparsity of the Hamiltonian, it may also
improve the computational scaling for DFT on large systems. In particular, using wavelet localization to
impose locality only at the atomic level may be more efficient than existing wavelet approaches which do
not recombine the wavelets into a smaller number of functions.  Specifically, one could wavelet filter
a standard Gaussian basis to produce an orthogonal basis with more locality and sparsity 
than traditionally localized Gaussian bases. 

Since we are trying to produce basis sets for correlated calculations, where basis set convergence
is slower than for DFT or HF calculations, we must think
about the effect of the basis on the electron-electron
cusp.
Our choice of 1D potential interaction, which has a slope discontinuity at the origin, is designed to
partially mimic the electron-electron cusp behavior in 3D.  In 3D, the potential diverges as $r \to 0$, but
the effect is substantially reduced by the 3D volume element. The moderate singularity we have in 1D is
similar, but we cannot expect our results
to match 3D precisely. Also, when trying to achieve chemical accuracy,
the short range cusp behavior is thought to be less relevant than intermediate distance electron-electron
correlation.  This further complicates the comparisons between 1D and 3D, and a 3D procedure and
benchmark calculations are
clearly needed.

Another difficulty in implementing our approach in 3D is the computation of the
integrals defining the Hamiltonian, once the basis is defined.  In our 1D implementation, 
all integrals are written in terms of sums over the fine grid; this would not be practical in 3D.
Wavelet bases, which are a crucial part of our wavelet localization, are able to represent
nuclear cusps more efficiently than grids, so one might try to work directly in 
the wavelet basis, expressing all the final basis functions as linear combinations of
wavelet functions.\cite{harrison2004multiresolution,yanai2015multiresolution,harrison2016madness} 
However, wavelets are much less efficient than atom-centered Gaussians for representing
nuclear cusps, and so a much more efficient approach might be to try
to combine wavelets with a few Gaussians per nucleus. Another approach
to dealing with nuclear cusps would be to use pseudopotentials, so there are no cusps.  
Yet another is to employ a basis set that inherently has a
one dimensional 
structure.\cite{frediani2015real, losilla2012divide,parker2014communication,PhysRevLett.119.046401}
We leave this set of 3D implementation problems for future work.

\section{Acknowledgements}

This work was supported by the U.S. Department of Energy, Office of Science, 
Basic Energy Sciences under award \#DE-SC008696. 
T.E.B.~also thanks the gracious support of the Pat Beckman Memorial Scholarship
from the Orange County Chapter of the Achievement Rewards for College Scientists Foundation.  T.E.B.~graciously thanks Professor Filip Furche, Dr.~Shane Parker, Dr.~Vamsee K.~Voora, and Sree Balasubramani for their patience in explaining and introducing methods from quantum chemistry.  Access to and discussion about the Block code was provided to T.E.B.~by Professor Garnet K.~Chan and Professor Sandeep Sharma, both of whom we thank.

\label{page:end}
\bibliography{wcg,Master,oldwcg,massiveWCG}

\end{document}